\documentstyle[12pt]{article}
\begin{document}

\baselineskip=23pt

\begin{center}
{\Large\bf Symmetry Realization, Poisson Kernel\\
 and the AdS/CFT Correspondence}

\bigskip

\bigskip

Zhe Chang$^{a,b}$ ~ and~  Han-Ying Guo$^{a,c}$

\medskip
{\em
a. CCAST (World Lab.), P.O.Box 8730, Beijing 100080, China.\\

b. Institute of High Energy Physics, Academia Sinica,\\
P.O.Box 918(4), Beijing 100039, China.\footnote{Mailing address.
 Email: changz@hptc5.ihep.ac.cn. }\\

c. Institute of Theoretical Physics, Academia Sinica\\
P.O.Box 2735, Beijing 100080, China.\footnote{Mailing address.
 Email: hyguo@itp.ac.cn.}}

\end{center}

\bigskip
\bigskip
Two kinds of realizations of symmetry on classical domains or 
the Euclidean version of AdS space 
are used to study 
 AdS/CFT correspondence.
Mass of free particles is defined as an AdS group invariant, the Klein-Gordon and
Dirac equations for relativistic particles in the AdS space are set up as a
simple mimic in the case of Minkowskian space.
The bulk-boundary propagator on the AdS space
is given by the Poisson kernel. Theorems on the Poisson kernel
guarantee the existence and sole of the bulk-boundary propagator. The
propagator is used to calculate correlators of the theories that live on the
boundary of the AdS space and  show conformal invariance, which is desired
by the AdS/CFT correspondence.

 \bigskip\bigskip\bigskip \bigskip
{\bf 1.} To understand the relations between gauge field theories and string theory
is a longstanding problem. The answer to the question will offer us a theory
of strong coupling mysteries of QCD. The AdS/CFT correspondence states that
string theory in Anti-de Sitter (AdS) spacetime is holographically dual to
a conformal field theory (CFT) on the spacetime
boundary of the AdS\cite{01}--\cite{04}. Beyond reasonable
doubt, the Maldacena dualities\cite{01} between $4$-dimensional
${\cal N}=4$ supersymmetric $SU(N)$
gauge theory and the type IIB string theory on the background AdS$_5\times S^5$ has been
established. Equivalence between theories in different dimensions immediately
raises questions about how detailed bulk information in one theory can be
completely coded in low dimensional degrees of freedom\cite{05}--\cite{08}.
Although there exist
a large amount of evidence, we have not yet any direct translation of the
configurations of one theory to the other. Nonperturbative formulations for
string and M theory
should be set up to investigate such a translation between theories in
different dimensional spacetimes.

The Maldacena conjecture is heavily dependent on peculiar properties of the
AdS spacetime. It is natural to begin such a program from discussing
symmetry realizations of the AdS group in the AdS spacetime. Here we use
two kinds of realizations on classical domains\cite{09,10},
which may be viewed as the Euclidean version of the AdS spacetime,
to study some properties of the AdS/CFT correspondence.

{\bf 2.} It is useful to consider the $(n+1)$-dimensional AdS space
AdS$_{n+1}$ as a submanifold of a pseudo-Euclidean
$(n+2)$-dimensional embedding space with coordinates $(y^\alpha )=(y^0,y^1,\cdots,
y^{n+1})$ and metric
$$\eta_{\alpha\beta}={\rm diag}(+,\underbrace{-,-,\cdots,-}_{n},+)$$
with ``length squared''
$$(y^0)^2-\sum_{i=1}^n(y^i)^2+(y^{n+1})^2=a^2$$
preserved by the AdS group $SO(2,n)$ acting as
\begin{equation}
{y}^\alpha \rightarrow \tilde{y}^\alpha=\Lambda^\alpha_\beta y^\beta~,~~~~\Lambda^\alpha_\beta\in SO(2,n)~.
\end{equation}
The metric of the AdS in the embedding space is given by
\begin{equation}
ds^2=-\sum_{i=1}^n(dy^i)^2+(dy^0)^2+(dy^{n+1})^2~.
\end{equation}
Making use of the coordinate $y^\alpha$, we can discuss motions of free
relativistic particles in the AdS space.
For this aim, we first define the $(n+2)$-dimensional angular momentum
$L^{\alpha\beta}$ as
\begin{equation}
L^{\alpha\beta}=y^\alpha\frac{dy^\beta}{ds}-y^\beta\frac{dy^\alpha}
{ds}~.
\end{equation}
$L^{\alpha\beta}$ should be conservative for the free particle, so that
\begin{equation} \label{eom}
\frac{dL^{\alpha\beta}}{ds}=0
\end{equation}
should be its equations of motion.
On the other hand, $L^{\alpha\beta}$ may be viewed as a set of classical realizations for generators of the AdS group $SO(2,n)$.
At this representation, the Casimir-like invariant of the AdS group may be
 defined as square of mass of the free  particle as follows:
\begin{equation}\label{energy-mass}
\begin{array}{rcl}
m_0^2&=&\displaystyle\frac{1}{2a^2}L^{\alpha\beta}L_{\alpha\beta}\\
     &=&E^2-{\bf P}^2-\displaystyle\frac{{\bf L}^2}{a^2}~,
\end{array}
\end{equation}
where $\displaystyle E\equiv \frac{1}{a}L^{n+1,0}$ and $\displaystyle
{\bf P}\equiv \frac{1}{a} L^{n+1,i}$.

It is interesting to see that in the projective space realization (see below
in $4.$),
 we may identify $P^i$ and $L^{ij}$ as
momentum and angular momentum of the free particle in the AdS space,
respectively.  And Eq.(\ref{energy-mass}) gives rise to the
generalization of 
Einstein's famous formula $E^2-{\bf P}^2=m_{0E}^2$ in the special relativity.

On the other hand, it is well-known that the operator form of  Einstein's  formula
gives the Klein-Gordon equation -- the relativistic quantum equation of 
motion for a free scalar particle or the relativistic equation of motion for
free  scalar field. A simple mimic of quantization at flat spacetime gives
the operator counterpart of the classical variables $L_{\alpha\beta}$,
\begin{equation}
L_{\alpha\beta}\longrightarrow \hat{L}_{\alpha\beta}=i\left(y_\alpha
\frac{\partial}{\partial y^\beta}-
y_\beta\frac{\partial}{\partial y^\alpha}\right)~.
\end{equation}
The operator forms of $L_{\alpha\beta}$ are, in fact, generators of the AdS
group. The Corresponding Casimir operator is  of the form
\begin{equation}
\hat{Q}=\frac{1}{2a^2}\hat{L}_{\alpha\beta}\hat{L}^{\alpha\beta}~.
\end{equation}

At this stage, we can set up equations of motion in relativistic
quantum mechanics of AdS invariant as
eigenvalue equation of the Casimir operator $\hat{Q}$ and it is 
independent of details of representation of the AdS. The Klein-Gordon equation
for relativistic scalar particles in AdS spacetime can be introduced as
\begin{equation}
(\hat{Q}-m_0^2)\Phi=0~.
\end{equation}
In a general frame with coordinates $x^i$ and metric $g_{ij}$ ($i,j=0,~1,~
\cdots,~n$), the Casimir operator $\hat{Q}$ of the AdS group
is of the form
\begin{equation}
\hat{Q}=\frac{1}{\sqrt{-g}}\frac{\partial}{\partial x^i}\left(\sqrt{-g}g^{ij}
\frac{\partial}{\partial x^j}\right)~.
\end{equation}
Thus, a general form of the Klein-Gordon equation in AdS space is
\begin{equation}
\left[\frac{1}{\sqrt{-g}}\frac{\partial}{\partial x^i}\left(\sqrt{-g}g^{ij}
\frac{\partial}{\partial x^i}\right)-m_0^2\right]\Phi(x)=0~.
\end{equation}

For spinors, analogue to Dirac operator in the Minkowskian space, 
we can introduce an AdS invariant operator as
\begin{equation}
\hat{\cal L}=\frac{1}{2}\gamma^\alpha\gamma^\beta \hat{L}_{\alpha\beta}~,
\end{equation}
where
$$\{\gamma^\alpha,\gamma^\beta\}=2\eta^{\alpha\beta}~.$$
It is easy to check that
\begin{equation}
\frac{a^2}{2}\hat{L}^{\alpha\beta}\hat{L}_{\alpha\beta}+\frac{9}{4}a^{-2}=\left(\hat{\cal L}-\frac{3}{2}
a^{-1}\right)^2~.
\end{equation}
Therefore,  similar to the case of Minkowskian space, the Dirac operator in AdS
space is a square root of
the Casimir operator of the AdS group. Then the eigenvalue equation of 
the Dirac-like operator
$\hat{\cal L}$ may
be viewed as the relativistic equation of motion for spinors,
\begin{equation}
[-i\gamma^i(\partial_i-\Gamma_i)+m_0]\Psi(x)=0~,
\end{equation}
where $\Gamma_i$ is the Ricci rotational coefficient
$$[\Gamma_k,\gamma_i]=\partial_k\gamma_i-\left\{\begin{array}{c}
                                                l\\
                                                ik
                                                \end{array}\right\}\gamma_l~.$$

{\bf 3.} In what follows, let us focus on the Euclidean version of the AdS$_{n+1}$. 
If we write a real $(n+1)$-dimensional vector as $x=(x^0,x^1,\cdots,x^n)$ and
$x'$ is the transport of $x$, the AdS$_{n+1}$  can be expressed as

\begin{equation}
xx'<1~,
\end{equation}
with the metric
$$ds^2=-\frac{4a^2\displaystyle\sum_{i=0}^n(dx^i)^2}{(1-xx')^2}~.$$
The coordinates $x^i$ and $y^a$ are related as
\begin{equation}
\begin{array}{l}
y^0=\displaystyle a\frac{1+xx'}{1-xx'}~,~~~~~~iy^{n+1}=a\frac{2x^{n+1}}{1-xx'}~,\\
y^i=a\displaystyle\frac{2x^i}{1-xx'}~,~~~~~~(i=1,~2,~\cdots,~n)~.
\end{array}
\end{equation}
It is not difficult to show that the transformation
\begin{equation}\label{transform}
\chi=\frac{x-b-xx'b+x(2b'b-bb'I)}
{1-2bx'+bb'xx'}~,~~~~bb'<1~,
\end{equation}
makes the AdS ($xx'<1$) to AdS ($\chi\chi'<1$), boundary of the AdS
($xx'=1$)
to boundary of the AdS ($\chi\chi'=1$) and the center ($x=b$) of the AdS to the
center ($\chi=0$) of the AdS. Here $I$ denotes the identity operator in the
AdS.

The inverse transformation is of the form
\begin{equation}
x=\frac{\chi+b+\chi\chi'b+\chi(2b'b-bb'I)}
{1+2b\chi'+bb'\chi\chi'}~,~~~~bb'<1~.
\end{equation}
We see that, besides of the point $\chi=-\frac{b}{bb'}$, the
transformation (\ref{transform}) is 1-1 correspondent.

Besides the above  transformations, it is easy to see that
\begin{equation}\label{transform2}
\chi=x\Gamma~,~~~~~~\Gamma\Gamma'=I
\end{equation}
also transforms the AdS to AdS but with invariant original point.
The transformations (\ref{transform}) and (\ref{transform2}) give a kind of
realization of the AdS group on the AdS space. This realization may be called
hyper-sphere realization.

For this realization of the AdS group, there is an invariant differential of
the form on the AdS space
\begin{equation}
\frac{d\chi d\chi'}{(1-\chi\chi')^2}=\frac{dxdx'}{(1-xx')^2}~.
\end{equation}
And the invariant differential operator is\\ \\
$\displaystyle(1-\chi\chi')^{n+1}\sum_{i=0}^n\frac{\partial}{\partial \chi^i}
\left[(1-\chi\chi')^{1-n}
\frac{\partial \Phi(\chi)}{\partial \chi^i}\right] $
\begin{equation}\label{poisson}
\begin{array}{rcl}
&=&
\displaystyle(1-xx')^{n+1}\sum_{i=0}^n\frac{\partial}{\partial x^i}\left[(1-
xx')^{1-n}
\frac{\partial \Phi(x)}{\partial x^i}\right]\\
&=&0~.
\end{array}
\end{equation}
This is just the Laplace equation, i.e. the Klein-Gordon equation, of
the massless scalar field.

It is convenient to introduce the spherical coordinates
\begin{equation}
\begin{array}{l}
x=\rho u~,~~~~uu'=1~,\\
dxdx'=d\rho^2+\rho^2dudu'~,\\
dudu'=d^2\theta_1+\sin^2\theta_1d^2\theta_2+\sin^2\theta_1\sin^2\theta_2
d^2\theta_3+\cdots+\sin^2\theta_1\cdots\sin\theta_{n-1}d^2\theta_{n}~,\\
~~~~~~~~~~~~~~~~~~~~~~~~~~~~~~~~~~~~~~0\leq\theta_1,~\cdots,~\theta_{n-1}
\leq\pi~,~~~~0\leq\theta_{n}
              \leq 2\pi~.
\end{array}
\end{equation}
The unit area element on the AdS boundary is as follows:
\begin{equation}
d\Omega_u=\sin^{n-1}\theta_1\sin^{n-2}\theta_2\cdots\sin\theta_{n-1}d\theta_1
d\theta_2\cdots d\theta_{n}~.
\end{equation}
From
\begin{equation}
d\chi d\chi'=\left(\frac{1-bb'}{1-2bx'+bb'xx'}
\right)^2dxdx'~,
\end{equation}
we know that on the spherical surface $x=u,~\chi=v~,uu'=vv'=1$ there is a
relation
\begin{equation}
dvdv'=\left(\frac{1-bb'}{1-2bu'+bb'uu'}\right)^2dudu'~.
\end{equation}
For the  $n$-dimensional vector $d\Omega_u$, similarly, we have
\begin{equation}
d\Omega_v=\left(\frac{1-bb'}{1-2bu'+bb'}\right)^{n}d\Omega_u~.
\end{equation}
This suggests the Poisson formula
\begin{equation}\label{formula}
\begin{array}{rcl}
\Phi(x)&=&\displaystyle \frac{1}{\omega_{n}}\int_{uu'=1}\cdots\int
d\Omega_u G^E_{B\partial}(x,u)\Phi_0(u)\\
&=&\displaystyle\frac{1}{\omega_{n}}\int_{uu'=1}\cdots\int d\Omega_u
\left(\frac{1-xx'}{1-2xu'+xx'}\right)^{n}\Phi_0(u)~,
\end{array}
\end{equation}
where
\begin{equation}
G^E_{B\partial}(x,u)=\left(\frac{1-xx'}{1-2xu'+xx'}\right)^{n}
\end{equation}
is the bulk-boundary propagator in the AdS, $\omega_{n}\left(=\displaystyle
\int_{uu'=1}\cdots\int d\Omega_u=\frac{2\pi^{\frac{n+1}{2}}}{\Gamma\left(\frac{n+1}
{2}\right)}\right)$ is
the total areas of the AdS boundary in Euclidean version, $\Phi(x)$
satisfies Eq.(\ref{poisson}) and $\Phi(x)
\vert_{x\rightarrow u}=\Phi_0(u)$. Let $x=\rho v$, we can rewrite the
bulk-boundary propagator as
\begin{equation}
G^E_{B\partial}(x,u)=\left(\frac{1-\rho^2}{1-2\rho\cos\langle u,v\rangle+\rho^2}\right
)^{n}~,
\end{equation}
where $\langle u,v\rangle$ denotes the angle between the vectors $u$ and $v$.
The bulk-boundary propagator $G^E_{B\partial}(x,u)$ has the following
properties:
\begin{itemize}
\item Because $1-2\rho \cos\langle u,v\rangle+\rho^2\geq(1-\rho)^2$, we
always have $G^E_{B\partial}(x,u)>0~,~~~~$ for $0\leq\rho<1$.
\item \begin{equation}
      \lim_{\rho\rightarrow 1}G^E_{B\partial}(x,u)=\left\{\begin{array}{c}
                                           0~,~~~~{\rm for}~u\not=v~,\\
                                           \infty~,~~~~{\rm for}~u=v~.
                                           \end{array}\right.
      \end{equation}
\item \begin{equation}
      \frac{1}{\omega_{n}}\int_{uu'=1}\cdots\int d\Omega_u
      G^E_{B\partial}(x,u)=1~.
      \end{equation}
\item At the case of $\rho<1$, the bulk-boundary propagator satisfies Eq.(\ref{poisson}).
\end{itemize}
These properties guarantee the existence of the Poisson formula
Eq.(\ref{formula}).
This shows the existence of the bulk-boundary propagator for Klein-Gordon
equation. The AdS/CFT correspondence states that\cite{03}
\begin{equation}
Z_{\rm bulk}[\Phi(\Phi_0)]=\langle e^{-\int_{uu'=1}\cdots\int d\Omega_u
\Phi_0(u){\cal O}(u)}\rangle_{\rm CFT}~.
\end{equation}
Here $\ln Z_{\rm bulk}$ is the effective action for string theory on
AdS$_{n+1}$ considered as a functional of the boundary data $\Phi_0(u)$
for the
fields $\Phi(x)$. The right side is the generating functional of correlators of
the operator ${\cal O}(u)$ dual to $\Phi_0(u)$.
Based upon the above discussion, we
know that the free classical solutions of scalar particle may be
suggestively written in terms of
the boundary value $\Phi_0(u)$ and a bulk-boundary propagator $G^E_{B\partial}
(x,u)$.

Then the CFT correlators are given in terms of truncated bulk Green functions
as
\begin{equation}
\langle {\cal O}(u_1){\cal O}(u_2)\cdots{\cal O}(u_m)
\rangle_{\rm CFT}=
\int\prod_{i=1}^m[dx_iG^E_{B\partial}(x_i,u_i)]\langle \Phi(x_1)
\Phi(x_2)\cdots\Phi(x_m)\rangle_T~.
\end{equation}
To check our results, some sample calculations will now be carried out, in the
approximation of classical supergravity. Consider an AdS theory that contains
a massless scalar $\Phi$ with action
\begin{equation}
I(\Phi)=\frac{1}{2}\int_{xx'<1}\cdots\int d^{n+1}x\sqrt{g}g^{ij}
\frac{\partial\Phi}{\partial x^i}\frac{\partial\Phi}{\partial x^j}~.
\end{equation}
Using the bulk-boundary propagator $G^E_{B\partial}(x,u)$, the solution of
the Laplace equation in the AdS
space with boundary values $\Phi_0$ is
\begin{equation}
\Phi(x)=\int_{uu'=1}\cdots\int d\Omega_u \left(\frac{1-xx'}{1-2xu'+xx'}
\right)^n\Phi_0(u)~.
\end{equation}
It follows that
\begin{equation}
\frac{\partial\Phi(\rho,u)}{\partial \rho}=-2n(1-\rho^2)^{n-1}
\int_{vv'=1}\cdots\int d\Omega_v
\frac{\Phi_0(v)}{\left[(u-v)(u'-v')\right]^n}-{\cal O}\left((1-\rho^2)^{n+1}
\right)~,~~~~\rho\longrightarrow 1~.
\end{equation}
By integrating by parts, one can express $I(\Phi)$ as a surface integral,
\begin{equation}
I(\Phi)=(2a)^{n-1}n\int_{uu'=1}\cdots\int d\Omega_u\int_{vv'=1}
       \cdots\int d\Omega_v
          \frac{\Phi_0(u)\Phi_0(v)}{\vert u-v\vert^{2n}}~.
\end{equation}
Hence the two points function of the operator ${\cal O}$ is proportional to
$\vert u-v\vert^{-2n}$, as expected for a field ${\cal O}$ of conformal
dimensional $n$.

{\bf 4.} We have discussed the Euclidean version of AdS/CFT correspondence 
from symmetry realization point of view and in terms of the Poisson formula
for the hyper-sphere realization.
It is important to mention that  there is another realization of the symmetry
transformations of the AdS group, besides of (\ref{transform}). Namely, the 
real projective space realization of the AdS space and the transformations 
of the AdS group on it.

In this realization, the global transformations are given by
\begin{equation}
x \rightarrow \chi=\frac{\sqrt{1-bb'}(x-b)(1+\lambda b'b)}{1-bx'}~,
\end{equation}
where $bb'<1$ and
$$\lambda=\frac{1-\sqrt{1-bb'}}{bb'\sqrt{1-bb'}}~,~~~~~~
1+\lambda bb'=\frac{1}{\sqrt{1-bb'}}~.$$
The invariant differential metric for this kind of transformations is
of the form
\begin{equation}
\frac{d\chi(I-\chi'\chi)^{-1}d\chi'}{1-\chi\chi'}=
\frac{dx(I-x'x)^{-1}dx'}{1-xx'}~.
\end{equation}
At the boundary of the AdS space $x=u,~\chi=v$, we have $duu'=0$ and thus
\begin{equation}
dvdv'=\frac{1-bb'}{(1-bu')^2}dudu'~.
\end{equation}
The bulk-boundary propagator for this kind of symmetry is of the form
\begin{equation}
G^E_{B\partial}(x,u)=\frac{(1-xx')^{\frac{n}{2}}}{(1-ux')^{n}}~.
\end{equation}
The invariant differential of this kind of transformations,
which is satisfied by the above bulk-boundary propagator, is as follows:
\begin{equation}
\sum_{i=0}^n\frac{\partial^2\Phi}{\partial x^i\partial x^i}-\sum_{i,j=0}^n
x^ix^j
\frac{\partial^2\Phi}{\partial x^i\partial x^j}-2\sum_{i=0}^nx^i
\frac{\partial\Phi}{\partial x^i}=0~.
\end{equation}
At the boundary of the AdS space, the transformation law for unit area elements
is
\begin{equation}
d\Omega_v=\frac{(1-bb')^{\frac{n}{2}}}{(1-bu')^{n}}d\Omega_u~.
\end{equation}
After added of the transformations
\begin{equation}
\chi=x\Gamma~,~~~~~~\Gamma\Gamma'=I~,
\end{equation}
which making the original point ($x=0$) invariant, we complete this kind of
realization of the AdS group. Similar Euclidean version of the 
AdS/CFT correspondence with transformations
(\ref{transform}) can be presented also. 

What very interested here is that in this 
realization $x^i$ are what is called the Beltrami coordinates of the AdS.
For the non-Euclidean version, the properties of the Beltrami
coordinates of the AdS spacetime have been discussed thoroughly \cite{11, 12}.
It has been shown that for the AdS  spacetime
the Beltrami coordinates form an initial frame
 in certain sense. And the equations of motion (\ref{eom})
 for the classical free particle
just correspond to the first integral of the geodesic equation of the Beltrami
metric. In other words, the special relativity with AdS symmetry may be set up based
on the Beltrami coordinates. Solutions of the Klein-Gordon and Dirac equation
in the AdS space can also be given in a more straightforward way at the Beltrami
frame\cite{13}. 

Details about
these subjects will be presented in a forthcoming paper.
\vskip 2mm
\centerline{\bf Acknowledgments}

We would like to thank Prof. K. H. Look (Q. K. Lu) for enlightening discussions.
The work was supported in part by the National Science Foundation of China.


\begin{thebibliography}{99}
\bibitem{01} J. Maldacena, ``The large $N$ limit of superconformal field
             theories and supergravity'', Adv. Theor. Math. Phys. {\bf 2}
             (1998) 231, hep-th/9711200.
\bibitem{02} S. Gubser, I. Klebanov and A. Polyakov, ``Gauge theory
             correlators from non-critical string theory'', Phys. Lett.
             B{\bf 428} (1998) 105, hep-th/9802109.
\bibitem{03} E. Witten, ``Anti de Sitter space and holography'', Adv. Theor.
             Math. Phys. {\bf 2} (1998) 253, hep-th/9802150.
\bibitem{04} For a review, see, O. Aharony, S. Gubser, J. Maldacena, H. Ooguri
             and Y. Oz, ``Large $N$ field theories, string theory and
             gravity'', hep-th/9905111.
\bibitem{05} G. 't Hooft, ``Dimension reduction in quantum gravity'', in
             {\it Salamfest 1993}, P284, gr-qc/9310026.
\bibitem{06} L. Susskind, ``The world as a hologram'', J. Math. Phys.
             {\bf 36} (1995) 6377, hep-th/9409089.
\bibitem{07} L. Susskind and E. Witten, ``The holographic bound in Anti-de
             Sitter space'', SU-ITP-98-39, IASSNS-98-44, hep-th/9805114.
\bibitem{08} Z. Chang, ``Holography based on noncommutative geometry and the
             AdS/CFT correspondence'', hep-th/9904101, to be published in
             Phys. Rev. D.
\bibitem{09} L. K. Hua and K. H. Look, ``Theory of harmonic functions
             in classical domains'', Scientia Sinica {\bf 8} (1959) 1031.
\bibitem{10} L. K. Hua, ``Harmonic analysis for analytic functions of several
             complex variables on classical domains'' (in Chinese), Scientific
             Press (Beijing) (1958).
\bibitem{11} K. H. Look, Z. L. Zou and H. Y. Guo, ``The kinematic effects in the classical domains and the red-shift phenomena of extra-galactic objects, 
             Acta physica Sinica (In Chinese) 23 (1974) 225. 
\bibitem{12} Q. K. Lu, Z. L. Zou and H.Y. Guo, " Relativistic principle of 
             spacetime with constant curvature and its cosmological significance". 
             In the Proc. of the 3rd Marcel Grossmann Meeting on General 
             Relativity (1982), ed. by L. Hu. p97 and references therein.
\bibitem{13} G. Y. Li and H. Y. Guo, ``Relativistic quantum mechanics for
             scalar particles and spinor particles in the Beltrami-de Sitter
             spacetime'' (in Chinese), Acta Physica Sinica
             {\bf 31} (1982) 1501.

\end{thebibliography}
\end{document}